# Difficulty as a Proxy for Measuring Intrinsic Cognitive Load

Minghao Cai, Guher Gorgun, & Carrie Demmans Epp

EdTeKLA Research Group, Dept. of Computing Science, University of Alberta

minghao3@ualberta.ca, gorgun@leibniz-ipn.de, cdemmansepp@ualberta.ca

**Abstract**

Cognitive load is key to ensuring an optimal learning experience. However, measuring the cognitive load of educational tasks typically relies on self-report measures which has been criticized by researchers for being subjective. In this study, we investigated the feasibility of using item difficulty parameters as a proxy for measuring cognitive load in an online learning platform. Difficulty values that were derived using item-response theory were consistent with theories of how intrinsic and extraneous load contribute to cognitive load. This finding suggests that we can use item difficulty to represent intrinsic load when modelling cognitive load in learning games.

# Difficulty as a Proxy for Measuring Intrinsic Cognitive Load

## Objectives

The theory of cognitive load details the interactions between key cognitive processes, learning activity design, and student learning (Sweller, 1988). Similar to many educational constructs, cognitive load is a latent variable, thus we cannot directly observe the cognitive demands of an educational assessment, activity, or task and the cognitive load experienced by the learner. Typically, self-report instruments are used to measure cognitive load. When using self-reports, learners are asked to rate the degree of mental effort they exert for a given task, but it is difficult for them to identify levels of specific types of load. Learners may struggle to accurately reflect on an activity's cognitive load, especially its intrinsic load, making measurement unreliable. Consistent with this concern, De Leeuw and Mayer (2008) found that participants' ratings may reflect other cognitive factors in combination with cognitive load. Additionally, self-report measures are typically administered at the beginning, at the end, or within consistent intervals (e.g., every five minutes) during a task (De Leeuw and Mayer, 2008), making it hard to assess the cognitive load of each subtask or question in an assessment. While research studies have shown that other proxies (e.g., learner affect) can be used for predicting cognitive load (Cai et al., 2024), a further exploration of viable alternatives to self-report measures is needed. In this study, we investigate the utility of using the difficulty index of a question as a proxy for intrinsic load in an online literacy learning system.

## Theoretical Framework

According to cognitive load theory, working-memory capacity is limited when dealing with novel information. When the total cognitive load exceeds the working memory capacity, performance and learning process will be harmed (Cowan, 2014; Miller, 1956; Young et al., 2014). Cognitive load theory posits a triadic relationship among intrinsic load (i.e., the cognitive effort required by a task), extraneous load (i.e., factors related to how a learning task is designed) and germane load (i.e., the cognitive resources that a learner can use) (Sweller, 2010). For an optimal learning experience, a balance between all three should be established.

Given the relevance of cognitive load theory to establishing optimal learning environments, researchers have aimed to understand mental effort exerted by learners. Commonly used self-report measures require learners to have a clear and univocal interpretation of *mental effort* and task *difficulty* and an ability to identify how much mental effort they have exerted (Ouwehand et al., 2021). As the main criticism of self-report measures is their being subjective (Ouwehand et al., 2021), alternative measures are needed to capture cognitive load more systematically and reliably. Given these challenges, an exploration of viable alternatives to self-report measures is needed. In this study, we investigated the difference between learners' self-assessed perceptions of intrinsic load when using a literacy learning system and the difficulty of questions they engaged with as estimated using established psychometric methods, i.e., item response theory (IRT). Another goal of the study was to investigate the practicality and feasibility of adopting item difficulty indices as proxies for estimating intrinsic cognitive load.

## Methods

### Data Source

This study relies on two datasets. One was collected from language learners and is called English language learner data. The other, the item-difficulty database, is historical data of learner interactions with an adaptive literacy game. The collection and use of these datasets underwent institutional research ethics review and approval processes.

*English Language Learner Data*

Following consent procedures, data were collected from 35 English learners who interacted with an online literacy game for a duration of approximately one hour. These participants had no previous familiarity with the game.

Participants were between 17 and 33 years old ($M = 24.1$). The languages first spoken at home varied: 24 spoke Chinese, 4 spoke Japanese, and 3 spoke Persian (Farsi). Additionally, Korean, Gujarati, French, and Spanish were each reported as the home language for one participant.

The game continually inferred students' reading abilities using their performance on literacy tasks and covertly adapted the difficulty of reading materials and their associated questions. The difficulty level was initially set to grade 8; it was then automatically adjusted based on learner performance.

Throughout the study session, participants' self-reported cognitive load was recorded at approximately 5-minute intervals, resulting in 319 samples. The cognitive-load instrument consisted of 10 items and used a Likert rating scale, ranging from 1 (Strongly Disagree) to 10 (Strongly Agree). This instrument provides measures for overall cognitive load and each of its subtypes: intrinsic load, extraneous load, and germane load.

Logs of the learner's interactions with the system were recorded. These included the specific questions that each learner was assigned and the answers the learner provided. A total 2,054 questions were answered by learners. These answers were spread across 663 unique questions.

*Item Difficulty Database*

The database was provided by our commercial partner who obtained parent consent for data use as part of the system licensing agreement. Data from 4,874 students was provided and included two sets of items: independent questions and questions with a reading passage. There were 3,040 unique independent items and 8,102 unique items with a reading passage. We only used data from 9,227 items because the usage patterns did not allow further analysis (see below).

Item difficulty was estimated using IRT. We used IRT because IRT places learner ability levels and item difficulty on the same continuum and is regarded as a less sample dependent method (Osterlind & Wang, 2017). IRT estimates item difficulty based on a probability function. We used the Rasch model: the item discrimination parameters were set to 1 and we estimated item difficulty ($b$) using learner responses (Andrich, 2005). In this approach, higher $b$-values indicate more difficult items. We removed items with fewer than 100 responses for IRT analysis to establish stable item difficulty parameters (Osterlind & Wang, 2017). Because items could have been attempted more than once by learners, we selected the learner's first attempt for each item during pre-processing. For the independent questions, this resulted in the removal of 1,425 items. So, 1,615 items were included for the final analysis. For the questions that included a reading passage, this process resulted 2,903 items for IRT analysis.

**Analysis**

We identified the questions that the English learners answered between administrations of the cognitive load questionnaire (i.e., a learning segment). We mapped the completed questions to those in the question difficulty database and calculated the average difficulty for each learning segment.

Statistical analyses were performed to better understand how item difficulties and reported cognitive load varied. As the measures were collected in different scales, we standardized all measures, so their values ranged from 0 to 1. Heatmaps were created to illustrate the average values of measures for each participant to explore potential patterns across learners.

The identifiers for the learners are listed along the bottom of the heatmap (with each column representing an individual learner).

## Results

Learners reported experiencing moderate cognitive load ($M = .40$, $SD = .27$). While learners reported moderate levels of intrinsic ($M = .30$, $SD = .45$) and extraneous ($M = .54$, $SD = .49$) load, they reported high levels of germane load ($M = .75$, $SD = .43$).

**Figure 1**

*Heatmap for Average Item Difficulty (Diff_IRT), Reported Intrinsic Load (IL), and Cognitive Load (CL)*

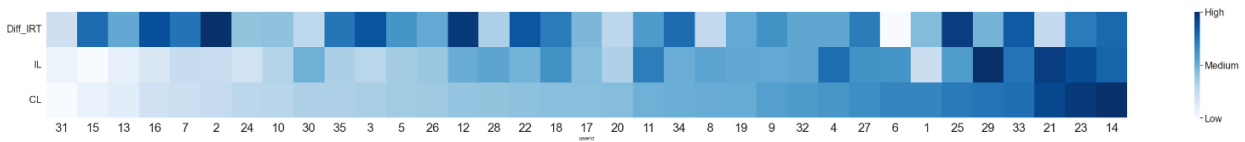

Figure 1 illustrates potential relationships among question difficulty, self-reported intrinsic load, and total cognitive load across learners, who are arranged by cognitive load score. The figure shows that self-reported intrinsic load mostly aligns with total cognitive load, which is expected since the intrinsic load scores contribute to the score for cognitive load (Sweller, 2011). This is also expected if learners were unable to separate their intrinsic load from their germane load when self-assessing (DeLeeuw & Mayer, 2008). Like intrinsic load, question difficulty was generally moderate ($M = .49$, $SD = .17$), and no clear relationship can be seen between question difficulty and cognitive load in Figure 1.

**Figure 2**

*Item Difficulty, Reported Intrinsic Load, and Cognitive Load over Time*

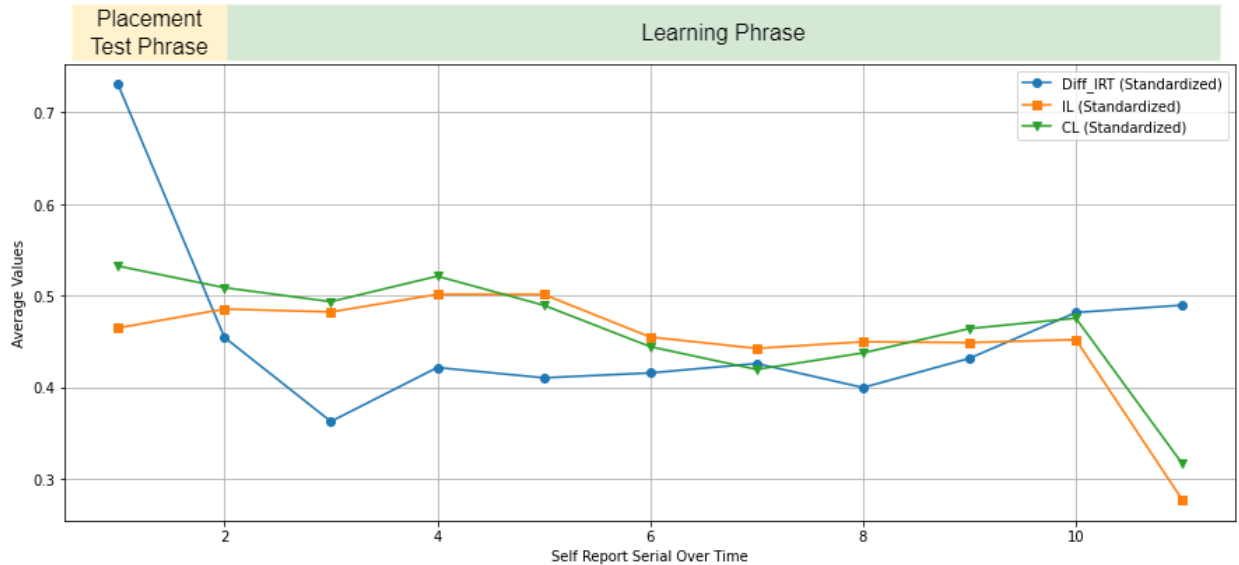

Next, we investigated these measures from a temporal perspective. Figure 2 shows item difficulty and self-reported learner intrinsic and cognitive load over time. The initial stage is the routing phase (i.e., item difficulty calibration), and the remaining time is the learning phase. In the routing stage, all learners were exposed to the same level of item difficulty and in the learning phase item difficulty was adjusted based on both the performance of the learner in the routing stage and their on-going performance. Thus, it is not surprising to find that in the initial phase, the item difficulty was high. It then dropped after the routing stage since the system adjusted the difficulty to better match learner performance. During the learning phase, the cognitive load measures were relatively stable until near the end of the session (Figure 2). This result indicates a discrepancy between students' perception of item difficulty and its actual level—it confirms the misalignment between learner perceptions of difficulty and difficulty as it is determined through learner performance.

**Figure 3**

*Heatmap for Sum of Item Difficulty and Reported Extraneous Load Comparing to the Reported Cognitive Load*

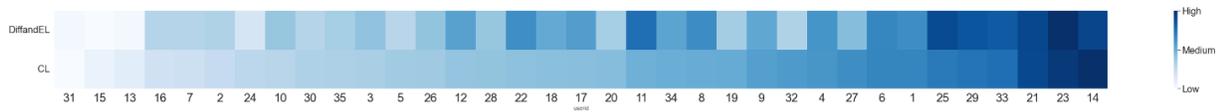

To investigate the feasibility of using question difficulty as a proxy for intrinsic load we summed the standardized extraneous load values and the average question difficulty to estimate overall cognitive load. We then used a heatmap to compare this sum with learner's self-reported cognitive load values. Figure 3 shows that the combination of difficulty index and extraneous load aligns with self-reported cognitive load.

**Figure 4**

*Trends of Sum of Item Difficulty and Reported Extraneous Load Comparing to the Reported Cognitive Load*

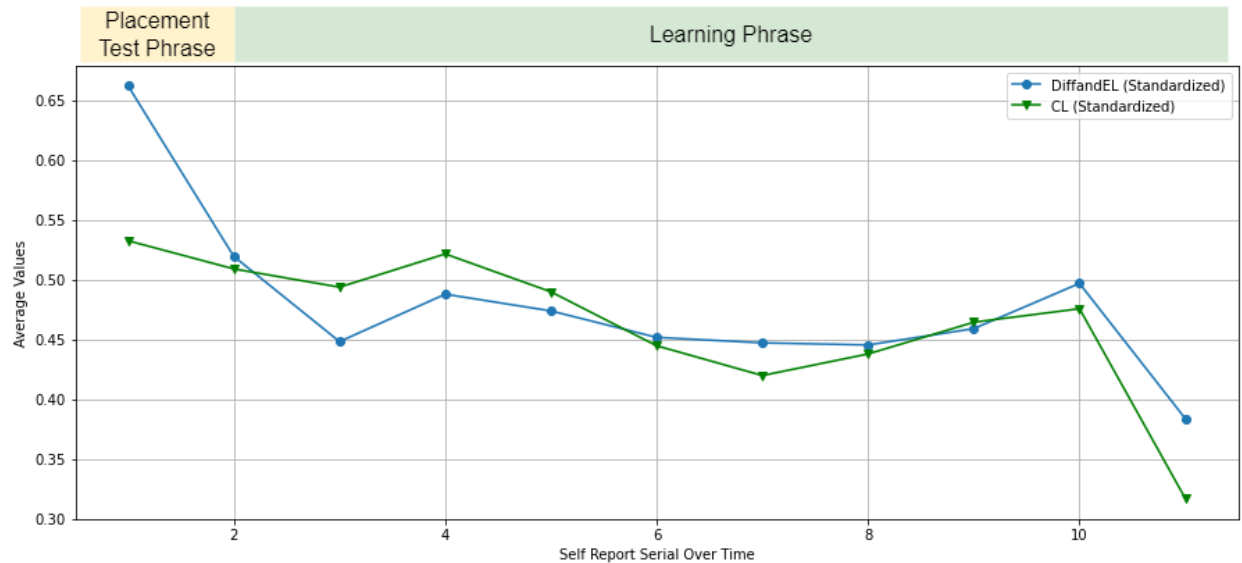

Additionally, from a temporal perspective (as shown in Figure 4), the changes in this combined value are consistent with those of self-reported cognitive load throughout the learning phase.

## Discussion and Significance

In cognitive load theory, intrinsic load reflects the load imposed by the difficulty of the learning task (Sweller, 2011). Our analyses indicate inconsistencies between learners' self-reported intrinsic load and the actual difficulty of the questions estimated by IRT using learner response data. While the difficulty of questions gradually increased in the learning phase, the measure of intrinsic load obtained via a questionnaire was generally decreasing. This incongruity may be due to learners' inability to separate intrinsic load from other types of load (DeLeeuw & Mayer, 2008) and their increased familiarity with the learning system. Additionally, learner reports of intrinsic load were lower during the routing phase despite the questions having higher difficulty than those encountered during the learning phase. These findings add to the body of literature challenging the reliance on self-report questionnaires (Ayres, 2008; Ouwehand et al., 2021), suggesting that self-report measures may not accurately reflect learner cognitive load. Findings further underscore the necessity for developing more reliable, data-driven alternatives for assessing cognitive load.

Our analyses indicate that combining the difficulty indices with the other forms of controllable load (i.e., extraneous load) creates values that are congruent with those of cognitive load. This finding is consistent with theories of how intrinsic and extraneous load contribute to cognitive load (Sweller, 2011) and suggests the feasibility of this approximation approach.

The implications of this approach are multifaceted. First, this method does not rely on self-reports, avoiding inaccuracies resulting from biases in self-perception and learner understanding of labels. It offers a more objective method of assessing cognitive load in educational settings, which could lead to more precise and reliable instructional design. Additionally, this method could enhance adaptive learning systems' efficiency by quantifying intrinsic load without interrupting the learners. This quantification could then be used to adapt learning content and optimize educational outcomes.

Our methodology underscores the need to reconsider the weight given to self-reported data in cognitive load research. While self-report measures are valuable for gaining insights into learners' subjective experiences, their limitations need to be acknowledged and supplemented with other data sources, such as interaction logs and measurement or inference approaches. We will continue to refine and explore the integration of this proxy measure to gain a more nuanced understanding of learner experience in complex educational settings. We will also investigate how adaptations based on estimated cognitive load may better support learning experiences and outcomes.

# References

Andrich, D. (2005). *The Rasch model explained. Applied Rasch measurement: A book of exemplars: Papers in honor of John P. Keeves* (pp. 27-59). Springer.

Cai, M., Rebolledo Mendez, G., Arevalo, G., Tang, S. S., Abdullah, Y. A., & Demmans Epp, C. (2024, May). Toward Supporting Adaptation: Exploring Affect's Role in Cognitive Load when Using a Literacy Game. *Proceedings of the CHI Conference on Human Factors in Computing Systems* (pp. 1-17).

Cowan, N. (2014). Working memory underpins cognitive development, learning, and education. *Educational Psychology Review*, *26*, 197-223.

DeLeeuw, K. E., & Mayer, R. E. (2008). A comparison of three measures of cognitive load: Evidence for separable measures of intrinsic, extraneous, and germane load. *Journal of Educational Psychology*, *100*(1), 223.

Miller, G. A. (1956). The magical number seven, plus or minus two: Some limits on our capacity for processing information. *Psychological Review*, *63*(2), 81.

Osterlind, S. J., & Wang, Z. (2017). Item response theory in measurement, assessment, and evaluation for higher education. In *Handbook on measurement, assessment, and evaluation in higher education* (pp. 191-200). Routledge.

Ouwehand, K., Kroef, A. V. D., Wong, J., & Paas, F. (2021). Measuring cognitive load: Are there more valid alternatives to Likert rating scales? *Frontiers in Education, 6*. 702616. https://doi.org/10.3389/feduc.2021.702616

Schmeck, A., Opfermann, M., Van Gog, T., Paas, F., & Leutner, D. (2015). Measuring cognitive load with subjective rating scales during problem solving: Differences between immediate and delayed ratings. *Instructional Science, 43*, 93-114.

Sweller, J. (1988). Cognitive load during problem solving: Effects on learning. *Cognitive Science*, *12*(2), 257-285.

Sweller, J. (2010). Element interactivity and intrinsic, extraneous, and germane cognitive load. *Educational Psychology Review*, *22*, 123-138.

Sweller, J. (2011). Cognitive load theory. In *Psychology of learning and motivation* (Vol. 55, pp. 37-76). Academic Press.

van Gog, T., Kirschner, F., Kester, L., & Paas, F. (2012). Timing and frequency of mental effort measurement: Evidence in favour of repeated measures. *Applied Cognitive Psychology, 26*, 833–839. https://doi.org/10.1002/acp.2883

Young, J. Q., Van Merrienboer, J., Durning, S., & Ten Cate, O. (2014). Cognitive load theory: implications for medical education: AMEE Guide No. 86. *Medical Teacher*, *36*(5), 371-384.